\documentclass[epj]{svjour}

\usepackage{graphicx}
\usepackage{amsmath,amsfonts,amssymb}

\begin{document}
\title{Scale-free memory model for multiagent reinforcement learning. Mean field approximation and rock-paper-scissors dynamics}
\titlerunning{Scale-free memory model for multiagent reinforcement learning}
\author{Ihor Lubashevsky\inst{1,2}\thanks{\,\email{ialub@fpl.gpi.ru}}
       \and
       Shigeru Kanemoto\inst{1}\thanks{\,\email{kanemoto@u-aizu.ac.jp}}
}                     
\institute{\mbox{University of Aizu, Tsuruga, Ikki-machi, Aizu-Wakamatsu City, Fukushima 965-8580, Japan}
    \and
\mbox{A.M. Prokhorov General Physics Institute, Russian Academy of Sciences, Vavilov Str. 38, Moscow 119991, Russia}
}
\date{Received: date / Revised version: date}
%
\abstract{
A continuous time model for multiagent systems governed by reinforcement learning with scale-free memory is developed. The agents are assumed to act independently of one another in optimizing their choice of possible actions via trial-and-error search. To gain awareness about the action value the agents accumulate in their memory the rewards obtained from taking a specific action at each moment of time. The contribution of the rewards in the past to the agent current perception of action value is described by an integral operator with a power-law kernel. Finally a fractional differential equation governing the system dynamics is obtained. The agents are considered to interact with one another implicitly via the reward of one agent depending on the choice of the other agents. The pairwise interaction model is adopted to describe this effect. As a specific example of systems with non-transitive interactions, a two agent and three agent systems of the rock-paper-scissors type are analyzed in detail, including the stability analysis and numerical simulation. Scale-free memory is demonstrated to cause complex dynamics of the systems at hand. In particular, it is shown that there can be simultaneously two modes of the system instability undergoing subcritical and supercritical bifurcation, with the latter one exhibiting anomalous oscillations with the amplitude and period growing with time. Besides, the instability onset via this supercritical mode may be regarded as ``altruism self-organization''. For the three agent system the instability dynamics is found to be rather irregular and can be composed of alternate fragments of oscillations different in their properties.
} 
\maketitle

\section{Introduction}\label{Intr}

During the last decades application of physical notions and the mathematical formalism of statistical physics to describing economic and social systems attracted much attention in the scientific community (see, e.g., \cite{ES2006}). The efficiency of this approach has been demonstrated, in particular, in modeling cooperative motion of vehicles in traffic flow, pedestrian ensembles, and groups of animals with social behavior \cite{Helbin}, dynamics of stock markets \cite{FM1,FM2}, opinion formation, culture and language evolution \cite{SD}. The multi\-agent reinforcement learning problem is one of the promising techniques of modeling the evolution and adaptation of complex systems in which human factor plays an essential role. Until now, this problem was studied mainly within the scope of artificial intelligence (for a review see \cite{TDAlg3}). Nevertheless, recently the concepts of statistical physics were combined with the notions of reinforcement learning to simulate the dynamics of minority games \cite{MG1,MG2,MG3,MG4} and evolutionary games \cite{EMG}, adaptive competition in a market \cite{ThM}, as well as to establish the relationship between the reinforcement learning and the replicator model of population biology \cite{RDRL1,RDRL2}. The latter, in particular, made it possible to analyze the complex behavior, including the onset of dynamical chaos for agent ensembles using the techniques of dynamical systems \cite{RD1,RD2,RD3}.

It should be noted that the majority of models similar to ones listed above and constructed to capture characteristic features of the social system dynamics invoke notions and concepts inherited directly from statistical physics. However, generally speaking, agent models imitating the behavior of human beings should possess also their own features that are inapplicable to physical objects or, at least, anomalous from the standpoints of physics \cite{Nat1,Nat2}. One of such anomalous features is the impact of system history implemented, in particular, via the effects of human memory in learning process as well as the evaluation of previous events in the decision making at the current moment of time. Comparing social systems and objects of classical physics we note that the latter ones have no memory in the following sense. If the positions and the velocities of all the particles forming a closed system are known at some moment of time chosen arbitrary, then its further dynamics is completely predictable at least  formally. Randomness arises at the level of reduced description and even in this case the dominating approximation is Markovian random processes again without memory; given the state of a system at some moment of time, its probabilistic dynamics is completely determined. Broadly speaking, in the present paper we intend to focus on possible history effects in social systems whose dynamics can be imitated by multiagent reinforcement learning as well as to discuss mathematical notions relevant to their description.

General arguments about human perception and event evaluation convince us to make use of a scale-free-memory model to describe the impact of system history on the reinforcement learning. In this model the impact of events happened in the past (at time $t'$) is weighted at the current moment of time $t$ with some function $K(t-t')$ showing a power-law decrease with the time difference $t-t'$. In fact, let us consider two events influencing our decision in a similar manner, which enables us to compare them with each other in assessing the current situation. If one of the two events happened one day before the current date whereas the other happened one week ago, then we will regard them as substantially different in time with respect to their contribution to our perception of the present situation. By contrast, if the first event occurred one month and one day ago and the second event occurred one month and one week ago, we will draw no real distinction between each other by the time of their occurrence. In other words, if the time lag between the two events is comparable with the time scale separating them from the present moment, then their impacts will be regarded as different in magnitude. On the contrary, if their time lag is much less then the passed time these events can be considered to be simultaneous. Exactly such a behavior is common to a power-law dependence $K(t-t')\propto (t-t')^{-z}$ with a certain exponent $z > 0$.

This model is partly justified by the observed long-term memory effects in the scale-free foraging by primates \cite{memory1,memory2,memory3} or insects \cite{memory4,memory5} and the conclusion about the explicit relationship between scale-free foraging and the memory properties \cite{memory6}. The human memory retrieval is also characterized by a scale-free pattern \cite{memory7} in addition to the fact that it is composed of many distinct systems (for a review see \cite{memoryAdd1}). Besides, stock markets, where human factor is doubtless of great concern, exhibit similar effects, in particular, time correlations in the volatility of returns are characterized by a power-law decay (see, e.g, \cite{Mark1,Mark2}).

The specific purpose of the present paper is to analyze how the scale-free memory can be introduced into the multiagent reinforcement learning as well as  to consider its characteristic effects in dynamics of multiagent systems with agent interaction similar to the rock-paper-scissors game. In such a system agents share a common environment, so each time, when one of them takes some action, it disturbs the environment, causing the response of the other agents. The learning process enables the agents with long-term memory to follow the variations of the environment in optimizing their own actions. The rock-paper-scissors model determining the agent interaction was chosen for two reasons.

First, this model is one of the simplest examples of non-transitive interactions: a rock beats a pair of scissors, scissors beat a sheet of paper, and paper beats a rock. So ordering the three objects according to their dominance we get a competitive cycle, which can be responsible for a variety of self-organization phenomena in social and biological systems (see, e.g., \cite{GamePhysics} and references therein).

Second, the rock-paper-scissors dynamics is found to be rather common for polymorphic populations, in particular, Jamaican cryptic coral reef communities \cite{RSPExam1}, yeast populations \cite{RSPExam2} and population of marine isopods \cite{RSPExam7,RSPExam8}, microbial laboratory communities \cite{RSPExam3,RSPExam4}, polymorphic groups of side-blotched lizards \cite{RSPExam5,RSPExam6}, lek-breeding ruffs \cite{RSPExam9,RSPExam10}, and Gouldian finches \cite{RSPExam11}. This fact merits a more detailed discussion.

\subsubsection*{Rock-paper-scissors game and speciation}

The trimorphism of such populations, wherein the species interaction is referred to as a biological rock-paper-scissors game \cite{RPSGame1}, is caused by basic properties of the frequency dependent selection being crucial in maintaining the biodiversity (for a review see \cite{FDS1}). The frequency dependent selection (FGS), i.e., the rate of offspring generation depending on the relative volume of a given phenotype, is implemented in such communities via various mechanisms. These mechanisms are classified into two groups, the positive and negative FDS, according to whether they cause this dependence to be increasing or decreasing, respectively. When mixtures of positive and negative FDS interact, a system can become destabilized and oscillate \cite{FDS1}.

Keeping in mind the further constructions, we note that learning and innate behavior are ones of the main mechanisms responsible for the positive FDS, with both of them affecting each other \cite{FDS1}. For example, the observed rapid evolution of dispersal, i.e., movement of individuals from their natal or previous breeding sites to new ones \cite{FDS2}, requires the presence of heritable genetic variations for traits affecting dispersal behavior and strong selection acting on these traits (see, e.g., a review \cite{FDS3}). In particular, in vertebrates, where dispersal is often considered to be a plastic, condition-dependent trait with low heritability, a significant heritability in the dispersal propensity, i.e. the probability of dispersal between habitat patches has been directly found for collared flycatchers \cite{FDS4}. Learning to be efficient requires enhanced long-term memory, especially it concerns those species of birds and mammals that cache food for later use or birds that carry out regular seasonal migrations returning to the same breeding and wintering grounds, or even stopover sites. For example, Clark's nutcrackers possess memory for cache sites spanning more than  7--9 months \cite{MB1} and for long-distance migratory garden warblers their memory of particular feeding sites persists during at least 12 months \cite{MB2}. It is worthwhile to note also investigations of the relationship between spatial use strategies and medial and dorsal cortical volumes in males of the side-blotched lizard as a criterion of the species memory capacity \cite{MB3}. Males in this species occur in three morphs using different spatial niches: large territory holders, small territory holders and nonterritory holders with home ranges of the smallest size.  According to the found results, the larger the used territory, the larger the medial and dorsal cortical volumes.

The analysis of the geographic variations and evolutionary history of the side-blotched lizard with the rock-paper-scissors mating strategies \cite{SP1} has thrown light on basic mechanisms governing speciation. Polymorphic forms within a population can be the starting material for new species \cite{SP2,SP3,SP4,SP5} and, in this case, the rock-paper-scissors FDS causing self-sustaining cycle variations maintains the required polymorphism \cite{SP1}. On one hand, the found pattern has turned out to contain also monomorphic and dimorphic populations that resemble the morphs in trimorphic populations. This resemblance suggests that rock-paper-scissors cycles of morphs were destabilized, giving rise to the irreversible loss and fixation of certain morphs and, thus, the rapid phenotype evolution.  On the other hand, the phylogenetic analysis revealed that this trimorphism was maintained for millions of years, even across speciation events. Thereby, the rock-paper-scissors interaction of species in trimorphic populations seems to cause evolutionary stable cyclic variations in morphs that undergo repeated breakdown events \cite{SP1}. In addition, analyzing a clade of dung beetles \cite{Beet} it has been demonstrated that insects assumed to be dimorphic can form trimorphic populations with cyclic phenotype variations via two threshold mechanism.

The model to be developed will demonstrate that long-term scale-free memory in the multiagent reinforcement learning governed by the rock-paper-scissors interaction can be responsible for complex dynamics. In particular, relatively regular oscillations can be alternated by oscillations with a large amplitude and period growing in time and the system admits the instability onset via subcritical as well as supercritical bifurcation for different modes of perturbations.

\section{Agent memory and reinforcement learning}\label{sec:mb}

\subsection{Continuous time description of multi-agent dynamics}\label{sec:2.1}

Let us consider of a collection of $N$ agents $\mathfrak{A}=\{a_i\}$ ($i=1,2,\ldots,N$)  that individually can take one of the actions from a set $\mathfrak{X} = \{x\}$  of $M$-elements and act \emph{independently} of one another. The preference of an action $x$ for a given agent $a$ is determined by the agent perception of its value $Q_a(x,t)$ gained by the current moment of time $t$ in exploring all the actions $\mathfrak{X}$ previously.

Within the discrete-time approximation the agents are assumed to take new actions at the time moments $t_n = n\,\Delta$, where $n=0,1,2,\ldots$ and $\Delta$ is the time step. The probability of choosing an action $x$ by a given agent $a$ at time $t$ is
\begin{equation}\label{mb:1}
     P_a(x,t) = e^{\beta Q_a(x,t)}\left[\sum_{x'\in \mathfrak{X}} e^{\beta Q_a(x',t)} \right]^{-1}\,,
\end{equation}
where the quantity $1/\beta$ characterizes the perception threshold of the agents in evaluating their actions.
If at the initial time $t_0=0$ (i.e. for $n=0$) the agents have no information about the action value, then the condition
\begin{subequations}\label{mb:2}
\begin{align}
\label{mb:2.0}
     \left.Q_a(x,t)\right|_{t=t_0} &= 0\\
\intertext{will hold for every agent $a$ and action $x$. Otherwise, the initial condition}
\label{mb:2.n0}
     \left.Q_a(x,t)\right|_{t=t_0} & = Q^*_a(x)
\end{align}
\end{subequations}
describes the agent preliminary opinion about the action value. In numerical simulations to be described below condition~\eqref{mb:2.n0} was used with the quantities $Q^*_a(x,t)$ set equal to some random numbers to disturb the system equilibrium and induce transient processes.

The system dynamics is governed by the learning of agents aimed at finding the most appropriate action via the trial-and-error search. Following \cite{RD1,RD2,RD3} we make use of a simple integrator algorithm of the reinforcement learning (see, e.g., \cite{TDAlg1,TDAlg2}). First, it assumes the agents to accumulate local rewards received at one step to raise awareness about the value of the possible actions. Second, because of bounded capacity of the agent memory events in the past separated from the present by time scales exceeding a certain value $T$ practically do not contribute to the awareness gained by the agents at the current time $t$. Third, according to expression~\eqref{mb:1} each agent explores more often actions in the vicinity of the action being optimal from its current point of view. So to reconstruct the value of the possible actions properly it should weight local rewards differently depending on the proximity of a given action to the optimal one. This reasoning is also supported by the fact that  people overweight low-probability events and underweight  high-probability events in the description-based decision-making which summaries descriptions detailing all possible outcomes and their respective likelihoods of each option (for a review see, e.g., \cite{DEG1,DEG2}). To allow for this feature we introduce the coefficient
\begin{equation}\label{app:2}
    W_a(x,t) = W\big[P_a(x,t)\big]
\end{equation}
weighting the reward gained by the given agent $a$ from taking the action $x$ at time $t$ depending on its current perception of the action value. Following the spirit of the prospect theory \cite{TK} and the update rule of frequency maximum $Q$-value heuristics~\cite{TDAlg3} the weight coefficient~\eqref{app:2} is assumed to be a non-increasing function on the current probability $P_a(x,t)$ of the agent $a$ taking the  action $x$.

The ansatz $W(P)=1$ is widely met in literature. However, when the $W(P)$-dependence is rather weak and the agent memory is long enough  the reinforcement learning mechanism under consideration can give rise to the Nash equilibrium instability not due to the agent interaction (Appendix~\ref{APP:MFA}). It is just caused by a higher rate of the reward accumulation for actions with higher probability. In the present paper we focus on the multiagent reinforcement learning whose complex dynamics is caused by the mutual interaction of agents possessing, in particular, long-term memory. So, from this standpoint such an instability may be treated as a ``malfunction'' of the analyzed algorithm and its possible influence on the studied phenomena should be eliminated. Keeping in mind results obtained in Appendix~\ref{APP:MFA} we adopt the ansatz
\begin{equation}\label{app:2sec}
    W_a(x,t) = \frac1{P_a(x,t)}
\end{equation}
which matches the simplest model being free from instabilities of the Nash equilibrium not caused by the agent interaction and describes the accumulation of knowledge about the action value proceeding uniformly, on the average, for all the possible actions.

Taking into account the aforementioned features the following version of the difference-learning equation
\begin{multline}\label{mb:3a}
     Q_a(x,t_{n+1}) = \frac{\delta_{xx_a}\Delta}{P_a(x,t_n)}\, R_a(x|\mathcal{X}_a)
\\
   {} - \frac{\Delta}{T}\,Q_a(x,t_{n})  + Q_a(x,t_{n})
\end{multline}
is applied to update the agent awareness at the time moments $\{t_n\}$. The first term in expression~\eqref{mb:3a} on the right-hand side describes the accumulation of the knowledge about the action $x_a$ taken by the agent $a$ at time $t_n$. Here $\delta_{xx_a}$ is the Kronecker delta, the payoff function $R_a(x|\mathcal{X}_a)$ describes the reward normalized to unit time that the agent $a=a_i$ gains from the action $x$ provided all the other agents
\begin{align*}
     \mathfrak{A}_a & = \{a_1, a_2, \ldots, a_{i-1}, \sqcup , a_{i+1},\ldots, a_N\}
\\
\intertext{have taken the actions}
     \mathcal{X}_a &= \{x_1, x_2,\ldots, x_{i-1},\sqcup, x_{i+1}, \ldots, x_N\} \,,
\end{align*}
and the cofactor $1/P_a(x,t)$ weights the contribution of the action $x$. The second term is caused by the agent memory loss and in what follows the inequality
\begin{equation}\label{mb:4}
     \Delta \ll T
\end{equation}
will be assumed to hold beforehand.

By virtue of inequality~\eqref{mb:4} for the function $Q_a(x,t)$ to reach some saturation a large number of the system updates need to be executed. In other words, every agent gains awareness of the action value through many trials and, thus, explores, explicitly or implicitly,  many options of the agent actions. It enables us, first, to average equation~\eqref{mb:3a} over all the possible points of the configuration space
\begin{equation*}
    \big\{\mathcal{X}\big\} = \{x_1, x_2,\ldots, x_N\}\,,\quad x_i\in \mathfrak{X}
\end{equation*}
assuming the probability of a particular configuration $\mathcal{X}$ occurring at time $t$ to be determined by the expression
\begin{equation}\label{PonChi}
    \mathcal{P}(\mathcal{X},t) = \prod_{i=1}^{N}P_i(x_i,t)
\end{equation}
and, second, to treat the action value $Q_a(x,t)$ as a continuous function of the time $t$. Decomposition~\eqref{PonChi} is due to the adopted assumption about the mutual independence of the agents in taking actions. In this way (Appendix~\ref{APP:MFA}) the update rule~\eqref{mb:3a} is reduced to the following differential equation
\begin{gather}\label{mb:5}
     \frac{dq_a(x,t)}{dt} =  r_a(x,t)- \frac{1}{T}\, q_a(x,t)\,,
\\ \intertext{where}
\label{mb:7}
     q_a(x,t) = Q_a(x,t) - \frac1M  \sum_{x'\in\mathfrak{X}} Q_a(x',t)
\end{gather}
and
\begin{multline}
\label{mb:6}
     r_a(x,t) = \sum_{\mathcal{X}_a} R_a(x|\mathcal{X}_a) \prod_{a'\in\mathfrak{A}_a} P_{a'}(x_{a'},t)
\\
     {} - \frac1M \sum_{x'\in\mathfrak{X}} \sum_{\mathcal{X}_a} R_a(x'|\mathcal{X}_a) \prod_{a'\in\mathfrak{A}_a} P_{a'}(x_{a'},t)
\end{multline}
is the reward rate gained by the agent $a$ from taking the action $x$  and measured relative to its value averaged over all the possible actions $\mathfrak{X}$.

In what follows we will confine our consideration to the dynamics of the quantities $q_a(x,t)$ instead of $Q_a(x,t)$ for two reasons. One of them is the fact that the probabilities $P_a(x,t)$ of the agent choice can equally be treated as direct functions of $q_a(x,t)$. Indeed, substituting \eqref{mb:7} into \eqref{mb:1} we get
\begin{equation}\label{mb:1q}
     P_a(x,t) = e^{\beta q_a(x,t)}\left[\sum_{x'\in \mathfrak{X}} e^{\beta q_a(x',t)} \right]^{-1}\,.
\end{equation}
The other is the possibility to eliminate a strong homogeneous growth of the action value from consideration which does not affect the system dynamics at all and, so, has no definite physical interpretation. The quantities $q_a(x,t)$ meet the equality
\begin{equation}\label{mb:7q}
    \sum_{x\in\mathfrak{X}} q_a(x,t) = 0\,.
\end{equation}
for any agent $a$ at each moment of time $t$ as follows form definition~\eqref{mb:7}.

It should be noted that expression~\eqref{mb:6} actually specifies some autonomous operator $r_a\big\{q_1,q_2,\ldots,q_N\big\}$ mapping the quantities $\{q_i\}$ onto the reward rate
\begin{equation}\label{mb:8}
     r_a(x,t) = r_a\big\{q_1(x,t),q_2(x,t),\ldots,q_N(x,t)\big\}\,,
\end{equation}
which holds also for a rather arbitrary form of the $W(P)$-dependence (Appendix~\ref{APP:MFA}). So, in fact, expression~\eqref{mb:5} is an autonomous nonlinear equation. It forms the complete continuous-time description of the multiagent system at hand provided the payoff function $R_a(x|\mathcal{X}_a)$ is known and the agent memory is characterized by the time scale $T$. To clarify the latter statement let us consider the integral representation of equation~\eqref{mb:5}.

\subsection{Memory models and the notion of initial conditions}\label{sec:MMIC}

Using the method of variation of parameters the differential equation~\eqref{mb:5} is reduced to the following Volterra integral equation
\begin{equation}\label{mb:9}
     q_a(x,t) = \int\limits_{t_0}^t dt'\, e^{-\tfrac{(t-t')}{T}} r_a(x,t')
     +
     e^{-\tfrac{(t-t_0)}{T}} q_a(x,t_0)\,,
\end{equation}
where the function $ r_a(x,t)$ is given by expression~\eqref{mb:8}. The Volterra equation~\eqref{mb:9} can be interpreted as the explicit formulation of the agent memory model characterized by the time scale $T$. The former term on the right-hand side of \eqref{mb:9} specifies the accumulation of the agent knowledge about the action value during the time interval $(t_0,t)$, whereas the latter one determines the evolution of the knowledge gained in the past. In fact, dealing with the whole history of the system we replace expression~\eqref{mb:9} by the corresponding integral over the semiaxis $(-\infty,t)$
\begin{align}
\label{mb:9a}
     q_a(x,t) & = \int\limits_{-\infty }^t dt'\, e^{-\tfrac{(t-t')}{T}} r_a(x,t')\,.\\
\intertext{Then using to the property of the exponential function}
\label{mb:9exp}
     e^{-\tfrac{(t-t')}{T}} & = e^{-\tfrac{(t-t_0)}{T}}\cdot e^{-\tfrac{(t_0-t')}{T}}\\
\intertext{we introduce the notion of initial conditions just setting}
\label{mb:9b}
     q_a(x,t_0) &= \int\limits_{-\infty }^{t_0} dt'\, e^{-\tfrac{(t_0-t')}{T}} r_a(x,t')\,,
\end{align}
converting \eqref{mb:9a} back to \eqref{mb:9}. In the frameworks of this model all similar events within time span about $T$ contribute to the agent perception equivalently and the function $K(t-t') = \exp\{-(t-t')/T\}$ is the kernel of the integral operator~\eqref{mb:9} weighting the current contributions of the events happened in the past.

If the agent memory is described by another kernel $K(t-t')$ not equal to the exponential function, i.e.
\begin{equation}
\label{mb:9K1}
     q_a(x,t) = \int\limits_{-\infty }^t dt'\, K(t-t')\, r_a(x,t')\,,
\end{equation}
then the property corresponding to equality~\eqref{mb:9exp} does not hold and the notion of initial conditions becomes inapplicable in the general case.

However appealing, for example, to ecological systems discussed in Introduction we see that the notion of initial conditions can have its own meaning independent of specific memory models. Habits of offspring being of multifactorial nature originate, in particular, from a mixture of parental and environmental contributions. The parental effects include a non-genetic parent-offspring transmission of preferable strategies of behavior reflecting the awareness gained previously by parents. The environmental factor is responsible for accumulating information of personal experience and its influence grows with increasing age. A discussion of these phenomena in connection with the biodiversity formation can be found, e.g., in reviews \cite{Plast1,Plast2,Plast3}. Therefore in mathematical description of the cumulative effect of natal awareness and personal experience on species adaptation the time of birth is the appropriate point for introducing the relevant initial conditions.

In the present paper we will construct initial conditions for the analyzed process of reinforcement learning presuming that there is a certain specific time moment $t_0$ when this process is initiated. Keeping in mind the parent-offspring communication let us adopt the following three assumptions about the learning process of agents with scale-free memory.

\emph{First}, within a sufficiently long time interval $T$ the agents remember the times $\{t'\}$ at which events happened and their contribution at the current moment of time $t$ is weighted by the kernel $K(t-t')\propto 1/(t-t')^{(1-\gamma)}$ with the exponent $0<\gamma<1$. The latter inequality is due the fact that, on one hand, the agent preference should be a really cumulative effect of all the previous rewards, i.e. the integral
\begin{equation}\label{memspec.1}
    C_{-}:=\int_{t-T}^t dt' K(t-t')\propto
    \int_{t-T}^t \frac{dt'}{(t-t')^{1-\gamma}}\sim \frac1\gamma\, T^{\gamma}
\end{equation}
has to diverge as formally $T\to \infty$. On the other hand, the kernel $K(t-t')$ must be a decreasing function. The estimate $C_{-}$ of integral~\eqref{memspec.1} can be regarded as a certain capacity of agent memory relating the action value $q_a(x)\sim \bar{r}_a(x)\, C_{-}$ to the mean rewards $\bar{r}_a(x)$ gained by the agent $a$ during this time.

\emph{Second}, on temporal scales larger than $T$ the agents do not rank the events according to their occurrence times, they just fix these events in the memory. It is described by the replacement
\begin{equation}\label{memspec.2}
     \int\limits_{-\infty }^{t-T} dt'\, K(t-t')\, r_a(x,t')\Rightarrow
      \bar{r}_a(x) \int\limits_{-\infty }^{t-T} dt'\, K(t-t') \,.
\end{equation}
So on such scales the integral
\begin{equation}\label{memspec.3}
   C_{+}\sim \int^{t-T}_{-\infty} dt' K(t-t')
\end{equation}
is to converge at the lower limit. In addition, its contribution to the memory capacity should be of the same order, i.e. the estimate $C_{+}\sim C_{-}$ must hold. The latter is the case if the kernel $K(t-t')\propto T^{2\gamma}/(t-t')^{1+\gamma}$ for $(t-t')\gtrsim T$. Here the factor $T^{2\gamma}$ is due to the continuity of the function $K(t-t')$ at $t-t'=T$.

\begin{figure}[t]
\begin{center}
\includegraphics[width=65mm]{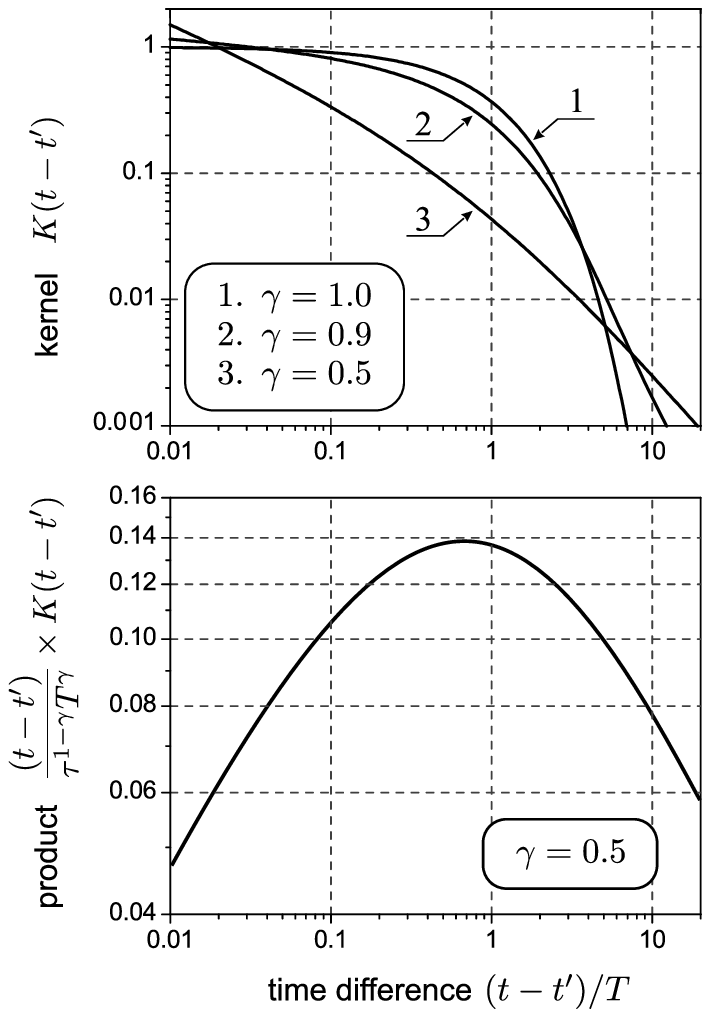}
\end{center}
\caption{A plot illustrating the excepted approximation~\protect\eqref{gammaexp}  of the kernel $K(t-t')$ vs the time difference $(t-t')$ (upper panel) and the crossover between asymptotics~\protect\eqref{asympK} (lower panel).}
\label{Fig1}
\end{figure}

Summarizing the two assumptions we state that the kernel $K(t-t')$ of the scale-free memory model should exhibit the following asymptotic behavior
\begin{subequations}\label{asympK}
\begin{align}
\label{asympK.a}
    K(t-t') \sim \frac{\tau^{1-\gamma}}{(t-t')^{1-\gamma}} & &\text{for $t-t'\lesssim T$}\,,\\
\intertext{and}
\label{asympK.b}
    K(t-t') \sim \frac{\tau^{1-\gamma}\,T^{2\gamma}}{(t-t')^{1+\gamma}} & &\text{for $t-t'\gtrsim T$}\,.
\end{align}
\end{subequations}
Here a certain ``microscopic'' time scale $\tau$ has been introduced because the kernel $K(t-t')$ must be a dimensionless quantity in the present constructions. Let us make use of the so-called $\gamma$-exponential function \cite{Kilbas} or, what is the same,  Rabotnov's function \cite{Podlubny}
\begin{equation}\label{gammaexp}
   K(t-t') =
    \frac{\tau^{1-\gamma}}{(t-t')^{1-\gamma}} E_{\gamma,\gamma}\left[-\left(\frac{t-t'}{T}\right)^\gamma \right]
\end{equation}
to construct the crossover between the given asymptotics. Here
\begin{equation}\label{MitLefE}
    E_{\gamma,\gamma}(z)
    = \sum_{k=0}^\infty\frac{ z^{k}}{\Gamma[(k+1)\gamma]}\,,
\end{equation}
is the Mittag-Leffler functions in two parameters and $\Gamma(z)$ is the gamma function. In the limit $\gamma\to1$, when all the events within the time scale $T$ contribute equivalently to the agent perception at the current time, kernel~\eqref{gammaexp} takes the exponential form, $K(t-t')\to \exp[-(t-t')/T]$. Figure~\ref{Fig1} illustrates the behavior of kernel~\eqref{gammaexp}.

\emph{Third}, at the initial time $t_0$ the agents have no personal knowledge about the value of the actions $\mathfrak{X}$  and can rely only on awareness gained previously in some way. It is a certain analogy to the situation described within the second assumption; only the fact that some event happened is essential, whereas its occurrence time is not known. So to quantify such pre-awareness we can deal with only some quantities $q_a(x,t_0)$ aggregating this information as a whole. To make it possible to measure the contributions from the pre-awareness and the personal experience in the same units let us introduce an effective reward rate $\bar{r}_a(x)$ related to the quantity $q_a(x,t_0)$ by the expression
\begin{equation}\label{mb:bc1add1}
    q_a(x,t_0) = \bar{r}_a(x) \int\limits_{-\infty}^{t_0} dt'\, K(t_0-t')\,.
\end{equation}
Then as time goes on the contribution of the pre-awareness evolves as
\begin{equation}\label{mb:bc1add2}
    \bar{r}_a(x) \int\limits_{-\infty}^{t_0} dt'\, K(t-t') =  K_b(t-t_0)\, q_a(x,t_0)\,,
\end{equation}
where the function
\begin{equation}\label{mb:bc1}
    K_b(t-t_0) = \frac{\left[\int\limits_{-\infty}^{t_0} dt'\, K(t-t')\right]}
    {\left[\int\limits_{-\infty}^{t_0} dt'\, K(t_0-t')\right]}\,.
\end{equation}
Using the Mittag-Leffler function in one parameter $E_\gamma(z)$ defined via the series \cite{Kilbas}
\begin{equation}\label{mb:bc2}
    E_\gamma(z)  = \sum_{k=0}^{\infty} \frac{z^{k}}{\Gamma(\gamma k+1)}
\end{equation}
we can show directly that
\begin{equation*}
    \frac{d}{dt}\, E_\gamma\left[-\left(\frac{t}{T}\right)^{\gamma}\right]  = -\frac1{t^{1-\gamma}T^{\gamma}} E_{\gamma,\gamma}\left[-\left(\frac{t}{T}\right)^{\gamma}\right]\,,
\end{equation*}
thus,
\begin{equation*}
   K(t-t') =   (\tau^{1-\gamma} T^{\gamma})\,\frac{d}{dt'}\, E_\gamma\left[-\left(\frac{t-t'}{T}\right)^{\gamma}\right]\,.
\end{equation*}
Therefore formula~\eqref{mb:bc1} can be rewritten as
\begin{equation}\label{mb:bc1final}
    K_b(t-t_0) = E_\gamma\left[-\left(\frac{t-t_0}{T}\right)^{\gamma}\right]
\end{equation}
provided the kernel $K(t-t')$ is given by expression~\eqref{gammaexp}.

Combing together the three assumptions we write the desired integral Volterra equation governing the multiagent reinforcement learning with scale-free memory in the following form
\begin{multline}\label{mb:9final}
     q_a(x,t) = \tau^{1-\gamma}\int\limits_{t_0}^t dt'\,
     \frac{E_{\gamma,\gamma}\left[-\left(\tfrac{t-t'}{T}\right)^\gamma \right]}{(t-t')^{1-\gamma}}\, r_a(x,t')
\\
     {}+ E_\gamma\left[-\left(\frac{t-t_0}{T}\right)^{\gamma}\right] \,q_a(x,t_0)\,.
\end{multline}
As before, the former term in this equation specifies the accumulation of the agent knowledge about the action value gained via reinforcement learning, whereas the latter one describes the evolution of the contribution from the pre-awareness. It should be noted that relative mathematical constructions were discussed within the so-called temporal-difference algorithm of the reinforcement learning \cite{TDAlg2}.

Concluding the given section we underline once more that the introduced notion of initial conditions implies the existence of a certain special point in the agent ``life''; it is the moment when the agents start their activity for the first time and, thus, have no direct experience in taking these specific actions. In contrast, the memory model with a fixed time scale enables one to impose initial conditions on the system dynamics at any moment of time.

\subsection{Governing equation}

The obtained integral equation~\eqref{mb:9final} can be converted into a differential equation with fractional time derivative using  fractional calculus. Namely, the known relationship between the Cauchy type problems for fractional differential equations and the Volterra integral equations \cite{Kilbas} enables us to reduce \eqref{mb:9final} to the following equation
\begin{equation}\label{mb:11}
   {}^C\!\widehat{D}_{t_0}^\gamma q_a(x,t) = \tau^{1-\gamma} r_a(x,t) - \frac1{T^{1-\gamma}}\,q_a(x,t)\,,
\end{equation}
where the left-hand side is the Caputo fractional derivative of order $\gamma$ defined by the expression
\begin{equation}\label{mb11RL}
   {}^C\!\widehat{D}_{t_0}^\gamma q_a(x,t):=
   \frac1{\Gamma(1-\gamma)}\int\limits_{t_0}^t \frac{dt'}{(t-t')^{\gamma}} \frac{dq_a(x,t')}{dt'} \,.
\end{equation}
Equation~\eqref{mb:11} should be subjected to the initial condition~\eqref{mb:2} or, more strictly, to the condition
\begin{equation}\label{mb:11incon}
   q_a(x,t_0) = q_a^*(x)
\end{equation}
with the quantities $q_a^*(x) = Q^*_a(x) - \left<Q^*_a(x)\right>_x$ given beforehand. Expression~\eqref{mb:11} is the desired governing equation of the mean field theory for the analyzed multiagent reinforcement learning with scale-free memory.

We also point out that for the scale-free memory the description of the multiagent reinforcement learning is no longer reduced to the replicator equations of population biology. For this reduction to hold the governing equation of the reinforcement learning has to be of the first order in the time derivative.

\subsection{Pairwise agent interaction}\label{sec:PWI}

To complete the construction of the model at hand we need to specify the interaction of the agents which is determined by the payoff function $R_a(x|\mathcal{X}_a)$. Let us confine our consideration to the pairwise approximation of the agent interaction \cite{ctmodel4}. In particular, biodiversity effects can be largely described in terms of pairwise interactions among species \cite{MSE1,MSE2}, however, multi-species interactions seem to be necessary in describing complex hierarchical communities \cite{MSE3,MSE4}. In the model of pairwise agent interaction the payoff function $R_a(x|\mathcal{X}_a)$ is written as (cf., e.g., \cite{PayoffMatrix})
\begin{equation}\label{mb:15}
     R_a(x|\mathcal{X}_a) = \rho_a^{x} + \sum_{a'\in\mathfrak{A}_a}\rho_{aa'}^{xx'}\,.
\end{equation}
Keeping in mind formula~\eqref{mb:6} determining the reward rate $r_a(x,t)$ as well as the identity
\begin{equation*}
    \sum_{x\in\mathfrak{X}}P_a(x,t) = 1
\end{equation*}
it is worthwhile to rewrite expression~\eqref{mb:15} in such a way that eliminates the terms not contributing to $r_a(x,t)$ and combines similar terms with one another. Namely, let us make use of the following replacements
\begin{align*}
  \rho_a^x  &:=  \rho_a^x - \Big<\rho_a^y\Big>_y
            + \sum_{a'\neq a}\left[\Big<\rho_{aa'}^{xy'}\Big>_{y'} - \Big<\rho_{aa'}^{yy'}\Big>_{yy'} \right]
\\
 \intertext{and}
   \rho_{aa'}^{xx'} &:= \rho_{aa'}^{xx'} - \Big<\rho_{aa'}^{xy'}\Big>_{y'} -\Big<\rho_{aa'}^{yx'}\Big>_{y} + \left<\rho_{aa'}^{yy'}\right>_{yy'}\,,
\end{align*}
where the notations
\begin{equation*}
   \Big<\ldots\Big>_y = \frac1M \sum_{y\in\mathfrak{X}} (\ldots) \quad\text{and}\quad
   \Big<\ldots\Big>_{yy'} = \frac1{M^2} \sum_{y,y'\in\mathfrak{X}} (\ldots)
\end{equation*}
have been introduced. In this case expression~\eqref{mb:6} becomes
\begin{equation}\label{mb:16}
     r_a(x,t) = \rho_a^x + \sum_{a'\in\mathfrak{A}_a}\sum_{x'\in\mathfrak{X}}\rho_{aa'}^{xx'}P_{a'}(x',t)
\end{equation}
and without loss of generality in what follows we will presume the equalities
\begin{equation}\label{mb:17}
   \Big<\rho_{a}^{y}\Big>_{y} = 0\,,\quad
   \Big<\rho_{aa'}^{yx'}\Big>_{y} = 0\,,\quad
   \Big<\rho_{aa'}^{xy'}\Big>_{y'} = 0
\end{equation}
to hold beforehand.

\section{Rock–-Paper-–Scissors model}\label{sec:rps}

\begin{figure}[t]
\begin{center}
\includegraphics[width=30mm]{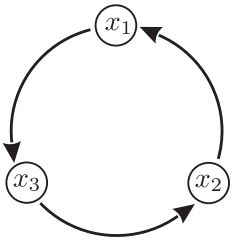}
\end{center}
\caption{Diagram illustrating the agent interaction of the rock-paper-scissors type.}
\label{Fig2}
\end{figure}

In the present section we consider a simple system of the rock-paper-scissors type and demonstrate that the scale-free memory can give rise to a complex dynamics caused by instability modes with different time scales. The specific model to be constructed mimics, in particular, some characteristic features of mating behavior in trimorphic populations of the side-blotched lizard (see \cite{Sex1} and references therein); a general review on the role of sexual selection in maintaining biodiversity can be found in \cite{Sex2}. In these populations female preferences impact on the frequency dependent selection and, what is essential, change with the female age. Females of the side-blotched lizard exist in two distinct morphs, however, in describing the female choice based on multiple male traits three types of females can be singled out \cite{3morph}. Males express three morphs that exhibit alternative strategies in intrasexual competition referred to as the rock-paper-scissors biological game (see Introduction). Recently, the mate choice governed by learning and accumulation of individual experience has attracted theoretical attention, in particular, when of the optimal choice is disturbed in a varying environment \cite{Sex3} or as a self-organization phenomenon in mutual male-female learning \cite{Sex4}; a review of the preceding models and used notions is also presented in the cited papers. In this context the model under consideration can be regarded as an example illustrating  the dynamics of female mate choice affected by the female intrasexual competition.

\subsection{Model}

Let us focus on two systems comprising two agents $\{a_1,a_2\}$ or three  agents $\{a_1,a_2,a_3\}$, and the set of their possible actions consisting of three elements $\{x_1,x_2, x_3\}$. All these actions on their own are considered to be equivalent for every agent, therefore, we set $\rho_a^x = 0$ for all $a$ and $x$. The implicit pairwise interaction of the agents with one another is determined by two factors. One of them is the structure of rewards gained by a pair of agents taking \emph{different} actions. These rewards are assumed to be determined by the rock-paper-scissors payoff matrix
\begin{equation*}
   \begin{bmatrix}
    0    &  1  & -1  \\
   -1    &  0  &  1  \\
    1    & -1  &  0
   \end{bmatrix}\,,
\end{equation*}
which is illustrated in Fig.~\ref{Fig2}. According to this payoff matrix if, for example, the agent $a_1$ takes the action $x_1$ and the agent $a_2$ takes the action $x_2$, then the first agent receives the benefit $g_{12}$, whereas the second one loses this value. The other factor is the reward redistribution when, for example, the agents $a_1$ and $a_2$ take the \emph{same} action $x_i$. To specify the latter factor we ascribe to each agent $a$ its individual ``power'' $0<\eta_a<1$ and assign, for example, to the agents $a_1$ and $a_2$ the rewards
\begin{equation}\label{rps:1b}
    \epsilon_{12} = g_{12}\omega_{12} \frac{\eta_1 }{\eta_1 + \eta_2}\,,\quad
    \epsilon_{21} = g_{21}\omega_{21} \frac{\eta_2}{\eta_1 + \eta_2}\,,
\end{equation}
respectively, with the interaction constants $g_{ii'}$ and $\omega_{ii'}>0$ being symmetrical with respect to permutation of indices.

Summarizing these assumptions and following the accepted renormalization of the payoff function $R_a(x|\mathcal{X}_a)$ (Sec.~\ref{sec:PWI}) we write the interaction matrix $\rho_{aa'}^{xx'}$ in the form
\begin{equation}\label{rps:2}
   \widehat{\rho}_{aa'}
   = g_{aa'} 
   \begin{bmatrix}
    \tfrac23\epsilon_{aa'}  &  1-\tfrac13\epsilon_{aa'}  & -1 -\tfrac13\epsilon_{aa'} \\
   -1-\tfrac13\epsilon_{aa'}  & \tfrac23\epsilon_{aa'}  &  1-\tfrac13\epsilon_{aa'} \\
    1-\tfrac13\epsilon_{aa'}  & -1-\tfrac13\epsilon_{aa'} &   \tfrac23\epsilon_{aa'}
   \end{bmatrix}\,.
\end{equation}
In addition the time scale $T$ characterizing the ability of the agent memory and introduced in Sec.~\ref{sec:MMIC} is set equal to infinity in order to study the effects of scale-free-memory on their own.

Under such conditions the governing equation~\eqref{mb:11} becomes
\begin{equation}\label{rps:3}
   \tau^{\gamma-1} {}^C\!\widehat{D}_{t_0}^\gamma q_a(x,t) =  \sum_{a'\in\mathfrak{A}_a}\frac{\sum_{x'}\rho_{aa'}^{xx'}e^{\beta q_{a'}(x',t)}}{\sum_{x''}e^{\beta q_{a'}(x'',t)}} \,.
\end{equation}
This equation possesses the stationary solution
\begin{equation}\label{Nash}
    q_a^\text{eq}(x) = 0
\end{equation}
for every agent $a$ and action $x$, which matches the Nash equilibrium attained when all the actions are equivalent in value and, thus, $P_a^\text{eq}(x) = 1/3$.

In what follows equation~\eqref{rps:3} will be analyzed with respect to the stability of the system dynamics and development of a possible instability will be studied numerically. In addition, to single out the affects that are due to the scale-free memory solely the further analysis will be confined to the case of identical agents setting $g_{aa'}=g$ and $\epsilon_{aa'}=\epsilon_{a'a}=\epsilon$ for any pair of agents $a$ and $a'$.

\subsection{Linear stability analysis}\label{Sec:LSA}

In the given multiagent system there are two sources of equilibrium perturbations. One of them is the deviation of the initial values $q^*_a(x)$ from $q_a^\text{eq}(x)$, the other is random variations in the reward rates reflecting the probabilistic nature of the agent choice. Since the initial conditions can be imposed on this system only at a certain specific moment of time the two types of perturbations need to be considered separately.

The present paper concerns the deterministic description of the agent choice dynamics and the equilibrium perturbations are introduced via the initial condition~\eqref{mb:11incon}. So in this section we explicitly analyze the system stability with respect to the first type perturbations, the second type perturbations are considered in Appendix~\ref{APP:B}.

The eigenfunctions of the Caputo fractional derivative operator~\eqref{mb11RL} meeting the Cauchy initial condition of type \eqref{mb:11incon} can be written in terms of the Mittag-Leffler function in one parameter, namely, $E_\gamma\left[\lambda (t-t_0)^{\gamma}\right]$, where $\lambda$ is the corresponding eigenvalue being a complex number in the general case. This follows from the identity \cite{Kilbas}
\begin{equation}\label{fexp}
     {}^C\!\widehat{D}_{t_0}^\gamma E_\gamma\left[\lambda (t-t_0)^{\gamma}\right] =
     \lambda E_\gamma\left[\lambda (t-t_0)^{\gamma}\right]\,.
\end{equation}
The known asymptotic behavior of the Mittag-Leffler function $E_\gamma(z)$ of order $0<\gamma<1$  \cite{Kilbas} enables us to represent the asymptotics of these eigenfunctions for $t\to\infty$ as
\begin{subequations}\label{asymp}
\begin{align}
\label{asymp1}
    E_\gamma\left(\lambda t^{\gamma}\right)  & = \frac{1}{\gamma}
     e^{\left(\lambda^{1/{\gamma}}\,t\right)}
     + O\left(\frac1{t^{\gamma}}\right)
\\
\intertext{when the argument of the eigenvalue $\lambda$ lies in the interval $|\text{arg}(\lambda)|\leq \gamma\pi/2$ and}
\label{asymp2}
    E_\gamma\left(\lambda t^{\gamma}\right)  & =
    - \frac1{\lambda\Gamma(1-\gamma)\cdot t^{\gamma}} + O\left(\frac1{t^{2\gamma}}\right)
\end{align}
\end{subequations}
for $\gamma\pi/2 <|\text{arg}(\lambda)|\leq \pi$. According to expressions~\eqref{asymp} the instability occurs when the governing equation~\eqref{rps:3} admits the eigenvalues meeting the inequality $|\text{arg}(\lambda)|\leq \gamma\pi/2$. In this case, by virtue of \eqref{asymp1}, the perturbation growth is exponential. The asymptotics~\eqref{asymp2} matching the stable system dynamics describes the power-law decay of perturbations.

Therefore linearizing equation~\eqref{rps:3} near the stationary point~\eqref{Nash} we seek its solution in the form
\begin{equation}\label{lin}
    q_a(x,t) = \theta_a^x E_\gamma\left[\lambda (t-t_0)^{\gamma}\right]\,,
\end{equation}
where $\{\theta_a^x\}$ are some constants. In this way the eigenvalue problem for equation~\eqref{rps:3} is reduced to finding the eigenvalues $h$ of the matrices
\begin{align}\label{matrix23}
    \widehat{\mathcal{F}}_2 &=
       \begin{bmatrix}
            0 & \widehat{\rho} \\
            \widehat{\rho} & 0
       \end{bmatrix}\,,
&
    \widehat{\mathcal{F}}_3 &=
       \begin{bmatrix}
            0 & \widehat{\rho} & \widehat{\rho}\\
            \widehat{\rho} & 0 & \widehat{\rho} \\
            \widehat{\rho}& \widehat{\rho}& 0
       \end{bmatrix}
\end{align}
for the systems with two and three agents, respectively, where the notation
\begin{equation}\label{rps:2simp}
   \widehat{\rho}
   =
   \begin{bmatrix}
    \tfrac23\epsilon    &  1-\tfrac13\epsilon  & -1 -\tfrac13\epsilon \\
   -1-\tfrac13\epsilon  & \tfrac23\epsilon     &  1-\tfrac13\epsilon   \\
    1-\tfrac13\epsilon  & -1-\tfrac13\epsilon  &   \tfrac23\epsilon
   \end{bmatrix}
\end{equation}
stands for matrices~\eqref{rps:2} in the case under consideration. The eigenvalues $h$ and $\lambda$ are related to each other via the expression
\begin{equation}\label{HandLambda}
    \lambda = \frac13\, hg\tau^{1-\gamma}\,.
\end{equation}
In addition, by virtue of \eqref{mb:7q} the corresponding eigenvectors are to meet the equality
\begin{equation}\label{eigenvectors:1}
    \sum_{x\in\mathfrak{X}}\theta_a^x(h) = 0
\end{equation}
for every agent $a$. Using Wolfram Mathematica~7 we have found the desired collection of the eigenvalues
\begin{subequations}\label{eigenvalues:12}
\begin{equation}\label{eigenvalues:1}
    \left\{-\epsilon \pm i\sqrt{3}\,, \epsilon \pm i\sqrt{3}\right\}
\end{equation}
for the two agent system and
\begin{equation}\label{eigenvalues:2}
    \left\{-\epsilon \pm i\sqrt{3}\,, 2(\epsilon \pm i\sqrt{3})\right\}
\end{equation}
\end{subequations}
for the three agent system, with the first two eigenvalues being doubly degenerate.

\begin{figure}[t]
\begin{center}
\includegraphics[width=65mm]{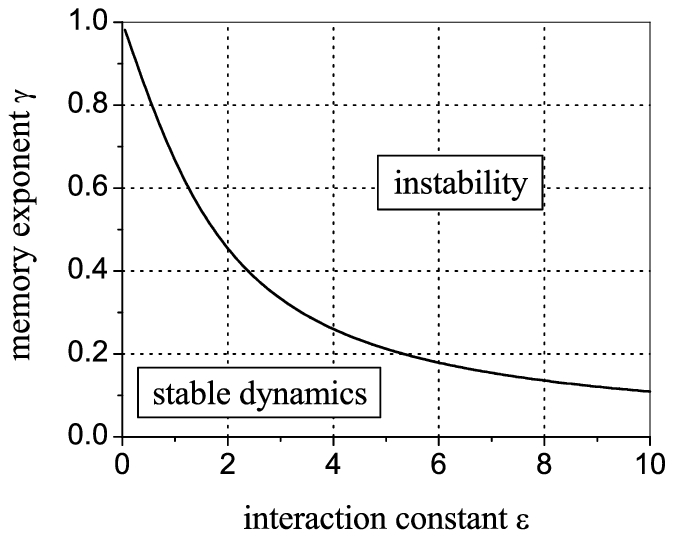}
\end{center}
\caption{Instability diagram for the analyzed systems with identical agents and $T\to\infty$.}
\label{Fig3}
\end{figure}

Whence it follows that both the two systems become unstable when $\text{arg}(\epsilon + i\sqrt{3}) < \gamma\pi/2$, i.e. the inequality
\begin{equation}\label{instab:1}
    \gamma > \frac{2}{\pi} \arctan\left(\frac{\sqrt{3}}{\epsilon}\right)
\end{equation}
holds. Figure~\ref{Fig3} depicts this condition.

As demonstrated in Appendix~\ref{APP:B} the found condition of the system instability holds also in the case when the equilibrium perturbations are caused by random fluctuations in the reward rates. So the complex system behavior found in the given deterministic model with dissipation cannot change drastically under the influence of small perturbations in the reward rates.

\subsection{Numerical simulation and the results}

\subsubsection{Algorithm}

Under the assumptions adopted in the previous section the governing equation~\eqref{rps:3} can be rewritten in the dimensionless form using the replacements
\begin{align*}
    t & \rightarrow \tau_p t\,, &
    \beta\, q_a(x,t) & \rightarrow  q_a(x,t)\,,
\end{align*}
where the characteristic time scale $\tau_p$ of the system dynamics is
\begin{equation}\label{tauP}
    \tau_p = \frac{\tau}{(\beta\rho\tau)^{1/\gamma}}\,.
\end{equation}
In these terms it takes the form
\begin{equation}\label{rps:3DL}
      {}^C\!\widehat{D}_{t_0}^\gamma\,\mathbf{q}_{2,3} = \widehat{\mathcal{F}}_{2,3}\, \mathbf{P}_{2,3}\,,
\end{equation}
where, for example, for the two agent system the vectors $\mathbf{q}_{2}$ and $\mathbf{P}_{2}$ denote the following collections of variables \begin{align*}
    \mathbf{q}_{2} &= \{q_1(x_1),q_1(x_2),q_1(x_3),q_2(x_1),q_2(x_2),q_2(x_3)\}
\\
    \mathbf{P}_{2} &= \{P_1(x_1),P_1(x_2),P_1(x_3),P_2(x_1),P_2(x_2),P_2(x_3)\}
\end{align*}
and the components of these vectors are related as
\begin{equation}\label{rps:3DLqp}
     P_a(x) =  e^{q_a(x)}\left[\sum_{x'=1}^3e^{q_{a}(x')}\right]^{-1}
\end{equation}
by virtue of \eqref{mb:1q}.

The right-hand side of equation~\eqref{rps:3DL} is a function of the vector $\mathbf{q}_{2,3}$ whose derivatives with respect to the components of this vector are bounded from above for any value of $\mathbf{q}_{2,3}$. It enabled us to make use of explicit algorithms in numerical simulation of the system dynamics (for discussion of this point see, e.g., \cite{Podlubny,Dain,Gaf,Galeone}). Namely, the governing equation~\eqref{rps:3DL} was solved numerically using the explicit 2-FLMM algorithm of second order in $\Delta$ \cite{Galeone} based on the following discretization of equation~\eqref{rps:3DL}
\begin{multline}\label{mb:12}
   \mathbf{q}_n = -\sum_{k=1}^{n-1} \omega^{(\gamma)}_{k}\mathbf{q}_{n-k} + b^{(\gamma)}_n \mathbf{q}_0
\\
   {} + \Delta^{\gamma} \widehat{\mathcal{F}} \left[\left(2-\frac{\gamma}{2}\right) \mathbf{P}(\mathbf{q}_{n-1})
       +\left(\frac{\gamma}{2}-1\right)  \mathbf{P}(\mathbf{q}_{n-2})
       \right].
\end{multline}
Here the indices denote the time moments $t_n = n\Delta$ ($n=2,3,\dots$) at which the corresponding quantities are taken, whereas the indices 2 or 3 labeling the systems under consideration are omitted for the sake of simplicity, $\{\omega^\gamma_k\}$ are the coefficients entering the Gr\"unwald-Letnikov definition of fractional derivatives specified, for example, via the following recursive formula
\begin{equation}\label{mb:13}
   \omega_0^{(\gamma)} = 1,\quad \omega_k^{(\gamma)} =\left(1-\frac{1+\gamma}{k}\right) \omega_{k-1}^{(\gamma)}
\end{equation}
for $k=1,2,\ldots$, and the coefficient
\begin{equation}\label{mb:13b}
   b_n^{(\gamma)} = \sum_{k=0}^{n-1}\omega_k^{(\gamma)}\,.
\end{equation}
The value $\mathbf{q}_1$ at the first step of the iteration was calculated as
\begin{equation}\label{mb:12.1}
   \mathbf{q}_1 = \mathbf{q}_0 + \Delta^{\gamma} \widehat{\mathcal{F}}\, \mathbf{P}(\mathbf{q}_0)
\end{equation}
and the initial value $\mathbf{q}_0$ meeting equality~\eqref{mb:7q} was set randomly to initiate system perturbations near the Nash equilibrium~\eqref{Nash}.

\subsubsection{Instability modes}

\begin{figure*}
\begin{center}
\includegraphics[width=130mm]{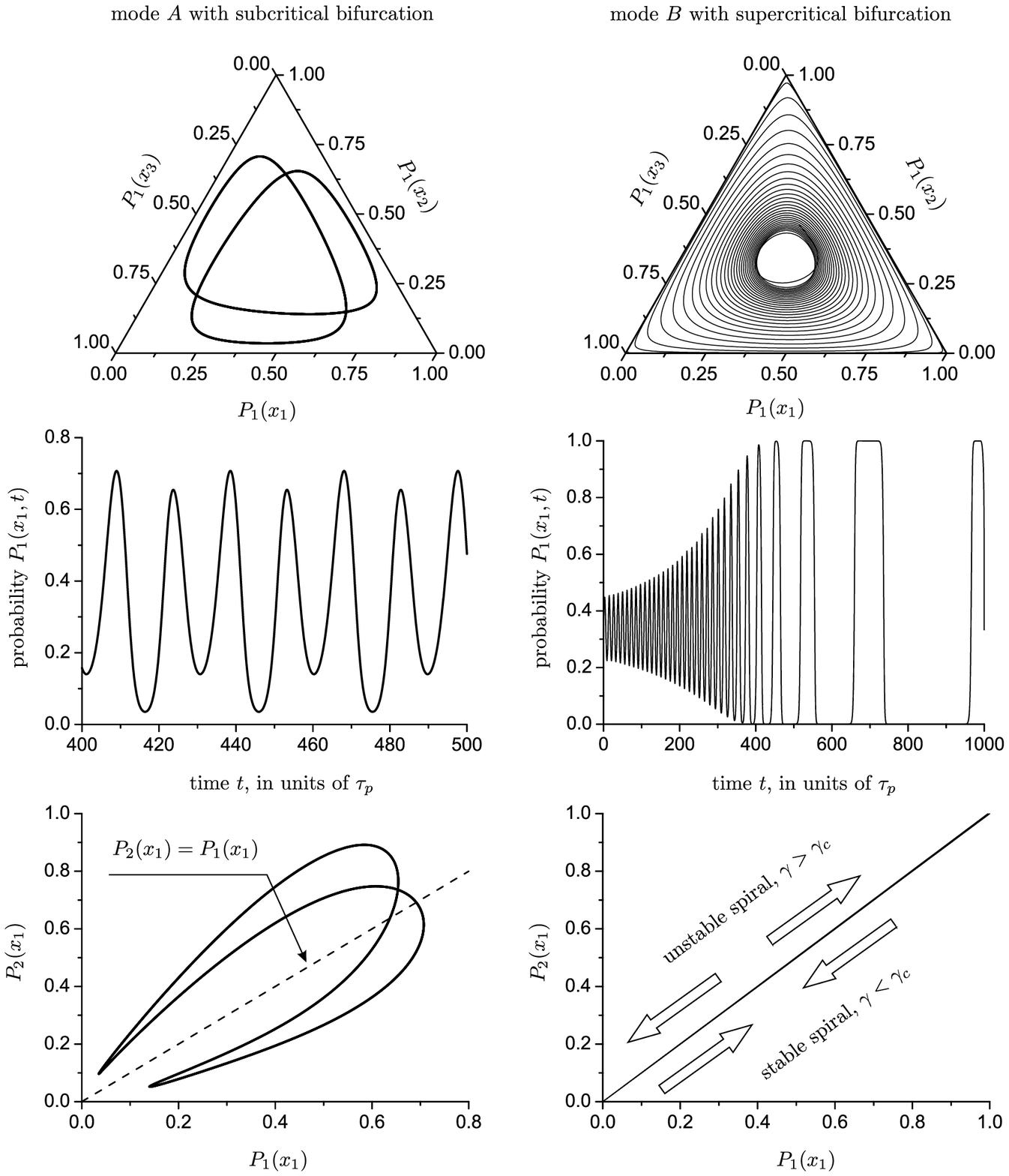}
\end{center}
\caption{Two modes of the long-term dynamics found for the two agent system. The upper row visualizes the dynamics of the agent $a_1$ as a ternary phase portrait of its trajectory $\{P_1(x_1,t),P_1(x_2,t),P_1(x_3,t)\}$  and the middle row shows the corresponding time pattern $P_1(x_1,t)$.  The agent $a_1$ and the action $x_1$ were chosen to exemplify the typical characteristics of the system dynamics. The lower row visualizes the correlations in actions of the two agents $a_1$ and $a_2$ in terms of the relationship between the probability of their choice of the action $x_1$. The present data were obtained for $\epsilon = 0.25$ and $\gamma = 0.91$. The critical value of the memory exponent is equal to $\gamma_c\approx 0.9087$ for the given magnitude of the parameter $\epsilon$.}
\label{Fig4}
\end{figure*}

\begin{figure*}
\begin{center}
\includegraphics[width=130mm]{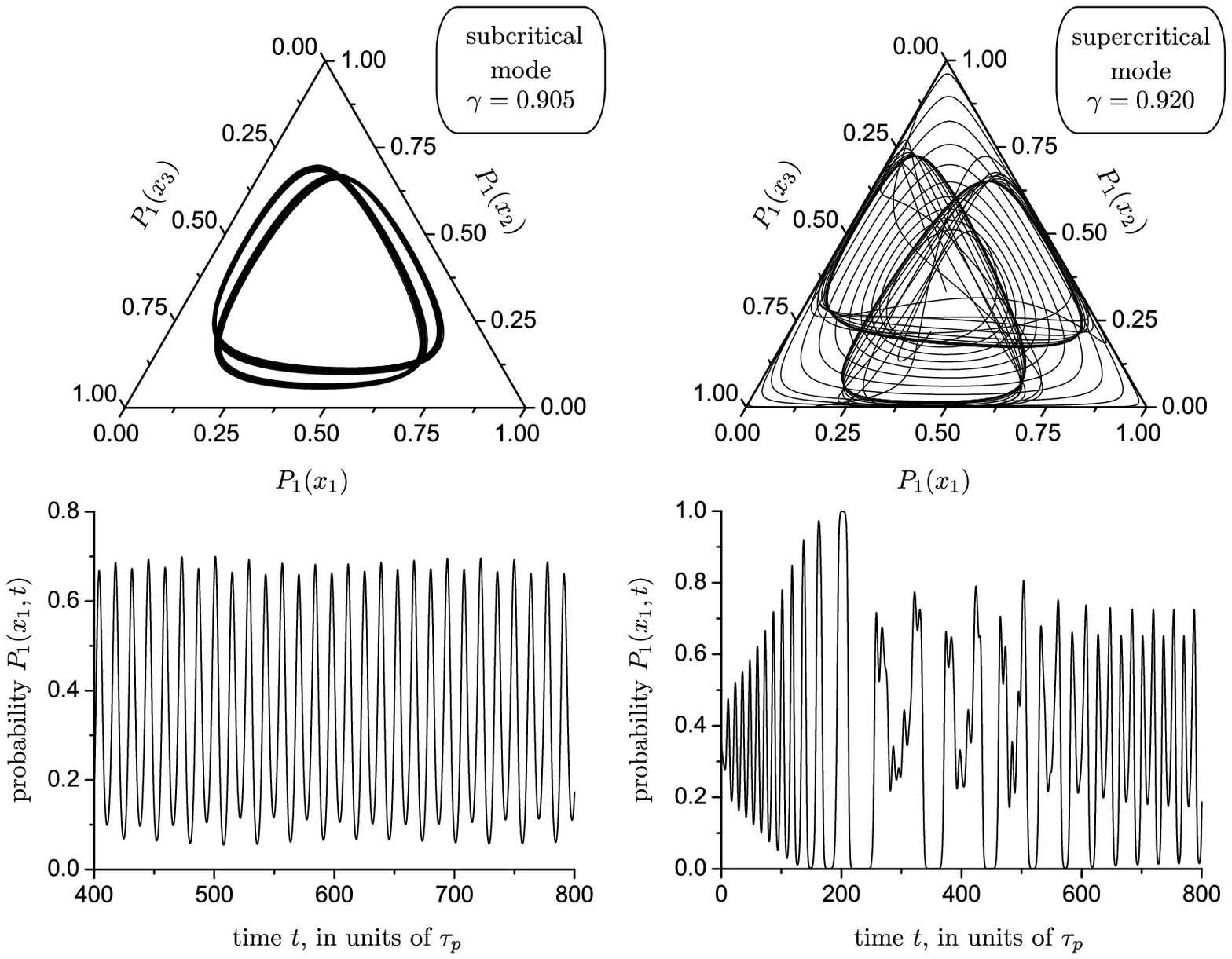}
\end{center}
\caption{The ternary phase portrait of a trajectory $\{P_1(x_1,t),P_1(x_2,t),P_1(x_3,t)\}$ and the corresponding time pattern $P_1(x_1,t)$ of the probability oscillations within the mode A in the subcritical and supercritical regions of the instability onset. The presented data were obtained for the two agent system with the parameter $\epsilon = 0.25$.}
\label{Fig5}
\end{figure*}

\begin{figure*}
\begin{center}
\includegraphics[width=130mm]{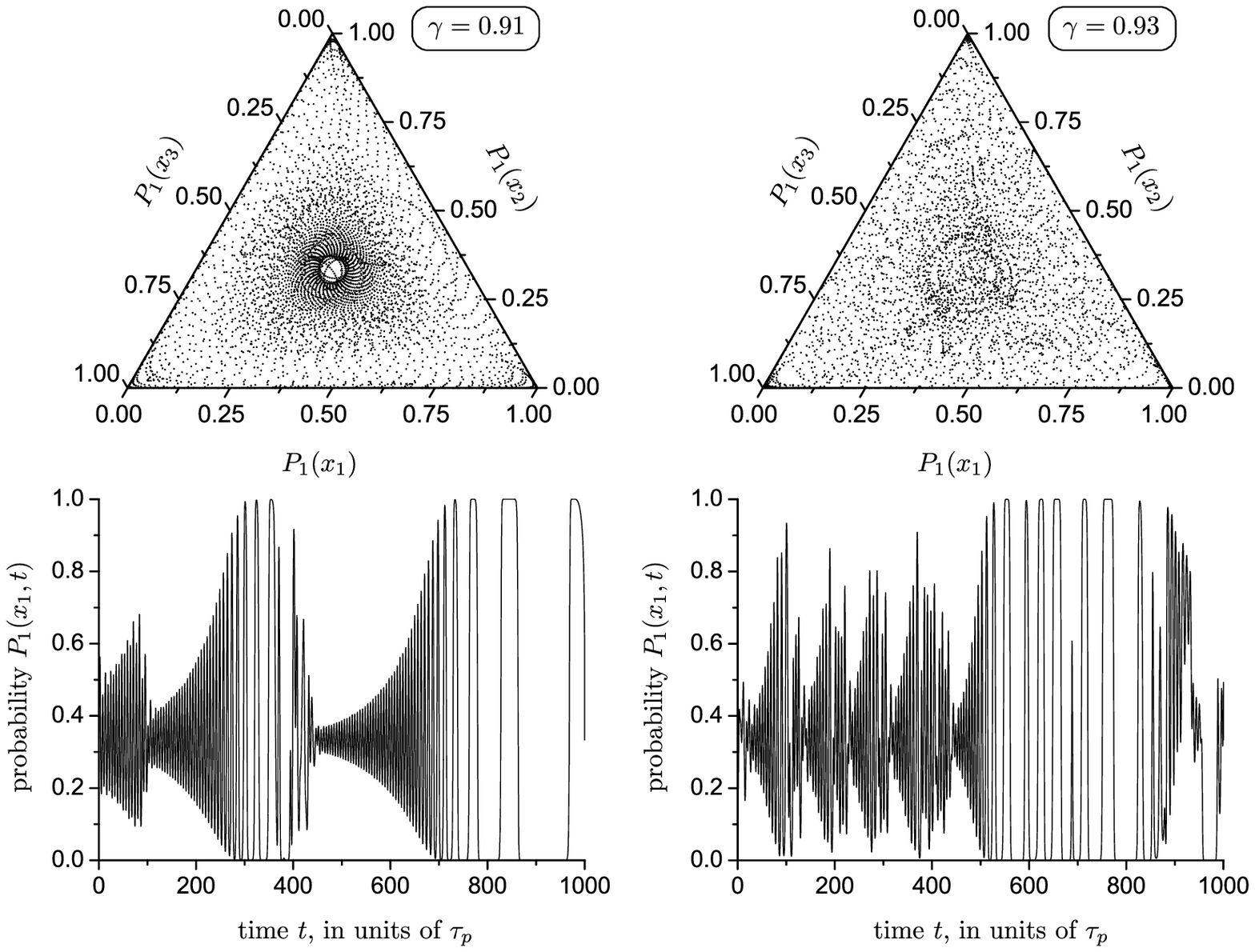}
\end{center}
\caption{The ternary phase portrait (in 6000 dots individually) of a trajectory $\{P_1(x_1,t),P_1(x_2,t),P_1(x_3,t)\}$ and the corresponding time pattern $P_1(x_1,t)$ visualizing typical features of the instability development for the tree agent system. The presented data were obtained for $\epsilon = 0.25$ and two values $\gamma = 0.91$ and $\gamma = 0.93$ of the memory exponent.}
\label{Fig6}
\end{figure*}

In the given paper we confine ourselves to discussing various modes of the system instability found numerically. Let us, first, present the results of simulation for the two agent system. Figure~\ref{Fig4} depicts two modes $A$ and $B$ of the long-term dynamics gotten by varying the initial conditions. The shown curves were obtained for the parameter $\epsilon = 0.25$ and the memory exponent $\gamma = 0.91$. On the stability diagram (Fig.~\ref{Fig3}) this point lies inside the instability region just near its boundary; for the given magnitude of the parameter $\epsilon$ the critical value of the memory exponent is $\gamma_c \approx 0.9087$.

The mode $A$ is related to a stable limit cycle in the phase space $\mathbf{P_1}=\{P_1(x_1),P_1(x_2),P_1(x_3)\}$ arising when a mismatch between the actions of the two agents is remarkable (Fig.~\ref{Fig4}, lower row). The mode $A$ was found can arise in the stability region also, i.e., when $\gamma < \gamma_c$, in particular, for $\gamma = 0.905$.  Figure~\ref{Fig3} characterizes the system stability only with respect to infinitesimal perturbations rather then perturbations with finite initial amplitudes. So, these results demonstrate us that the mode $A$ of the system instability undergoes the subcritical bifurcation as the memory exponent $\gamma$ increases. The periodic oscillations found in the subcritical region, $\gamma < \gamma_c$, are rather similar in form to those shown in Fig.~\ref{Fig4}. Only when the memory exponent $\gamma$ goes away from the critical value $\gamma_c(\epsilon)$ and comes close to a certain boundary of the absolute stability $\gamma_{s}(\epsilon)$ these oscillations exhibit more complex behavior. In particular, Fig.~\ref{Fig5} (left column) depicts steady state oscillations obtained for $\gamma = 0.905$. The corresponding trajectory of the system motion fills uniformly a certain neighborhood of the previous limit cycle. So such system motion and can be regarded as oscillations of the mode $A$ shown in Fig.~\ref{Fig4} whose amplitude undergoes some time variations. We have failed to find steady state oscillations for $\gamma = 0.904$ and $\epsilon = 0.25$; all the perturbations induced by initial random conditions faded out. It allows us to estimate the boundary of the absolute instability as $0.904<\gamma_s(\epsilon) < 0.905$ for the given value of the parameter $\epsilon$. So, when the mode $A$ of the system instability arises as the memory exponent $\gamma$ increases, its attractor seems to become rather complex in form instantly without a smooth transformation of a quasicircular line in the phase space  $\mathbf{P_1}$.

The second mode $B$ is related to the appearance of oscillatory instability undergoing the supercritical bifurcation, i.e. arising only in the instability region $\gamma>\gamma_c$. It turns out that even in the close proximity to the instability boundary the simplified model~\eqref{rps:3DL} seems not to be able to describe a possible steady state dynamics of the type $B$ instability. The found time pattern $P_1(x_1,t)$ exhibits the agent preference of taking only one action to become stronger and stronger as time goes on; the same does the duration of this choice (Fig.~\ref{Fig4}, right column). As also seen in this figure the mode $B$ matches the synchronized behavior of the two agents. It is likely that such oscillations can be stabilized on temporal scale of order $T$ by the capacity of the agent memory. In any case this feature is worthy of individual analysis. Now we can claim that, at least, the characteristic time scales of oscillations caused by the instability onset within the modes $A$ and $B$ differ from each other dramatically.

As the memory exponent $\gamma$ goes away from the instability boundary $\gamma_c$ inward the instability region the mode $A$ becomes dominant, whereas the mode $B$ looses its stability. It is demonstrated in Fig.~\ref{Fig5} (right column) visualizing an example of the transient processes of the instability development that at the initial stage can be classified as the mode $B$ and at the final stage converts into the mode $A$. Besides, the shown pattern being rather complex in structure enables us to presume that there can be other modes of the system instability which, at least, are metastable.

The found two modes $A$ and $B$ can be interpreted as a self-organization phenomenon of the ``selfish'' or ``altruistic'' behavior of agents; the term ``self-organization'' is used because in the given model the agents act independently of one another.  Indeed, within the mode $A$ each of the two agents alternately plays the role of defector  taking mostly the action ``beating'' the action chosen by the other agent. It is clearly demonstrated Fig.~\ref{Fig4} (lower row panels). Oppositely, within the mode $B$ the agents cooperate with each other taking mainly the same action. This behavior is really altruistic because the reward gained by the defector exceeds the reward gotten in sharing the same action by the factor $2/\epsilon$ for the used system parameters. The ``altruism self-organization'' seems to be due to the scale-free memory  because we have failed to find the mode $B$ in simulating actually the same system where, however, the agent memory is characterized by a single temporal scale (Sec.~\ref{sec:MMIC}). Up to now possible mechanisms responsible for the altruistic behavior are far from being understood well; a review of the corresponding phenomena and the recently proposed models can be found in \cite{Alt1,Alt2}. At least, altruism does not readily evolve as illustrated by evolution of cooperation in the prisoner's dilemma game \cite{Alt3}. In the given model the altruistic behavior matches the agent choice of one action for longer and longer time during the learning process. We note that similar dynamics was found in modeling cyclic variations in cooperator-cheat systems, where altruism and conspicuous tags of altruists are inherited via different mechanisms \cite{Alt4}.

Figure~\ref{Fig6} depicts typical details of the instability development for the three agent system. In this case the interaction between the agents destroys the mode $A$ and the system dynamics becomes irregular. However, as seen in this figure, near the instability boundary $\gamma_c \approx 0.9087$ for $\epsilon=0.25$ the mode $B$, nevertheless, can survive and emerge after a sufficiently long transient process during which the instability development is repeated several times. As a result the corresponding phase portrait in the form of dots exhibits some attraction of the system dynamics towards the origin visualized as a certain origin neighborhood of dot accumulation. Outside this narrow neighborhood of the instability boundary the system dynamics becomes more irregular. The corresponding phase portrait shown for $\gamma = 0.93$ matches a rather uniform dot distribution over the phase space. The related time pattern, nevertheless, again demonstrates the fact that the system from time to time returns to the origin and remains in its vicinity during a time interval determined by the instability increment. These results for the three agent system enable us to claim that the proposed multiagent model of the reinforcement learning with scale-free memory describes some anomalous mechanism forcing the system to return periodically to the origin and reside in its neighborhood during a remarkable time in spite of the origin being the unstable point. This mechanism could be a possible explanation for the observed evolutionary stable cyclic variations in trimorphic populations altered regularly by breakdowns and the formation of dimorphic or monomorphic populations (see Introduction).

\section{Conclusion}

A model for multiagent reinforcement learning with scale-free memory has been constructed. Its development was partly stimulated by our attempt to elucidate a way of describing memory effects in systems where living beings play a crucial role and which can be responsible for long-term phenomena observed in stock markets, scale-free foraging etc. The model details, including the agent non-transitive interaction of the rock-paper-scissors type, reflect characteristic features of trimorphic populations attracted much attention during the last decade in the context of speciation.

The reinforcement learning model assumes that the agents accumulate their rewards gained from taking some actions to get awareness about their value and, in addition, act independently of one another. The interaction arises implicitly  via the reward of one agent depending on actions of the other agents. The probability of taking an action is related to the gained awareness via an exponential function (Boltzmann model). The reward accumulation is described by an integral operator with a power-law kernel. In mean field approximation the final governing equation with fractional time derivative is constructed. The scale-free memory poses a question on the notion of initial conditions for such systems and a certain approximation allowing it has been found. Its key point is to relate the initial conditions with the time moment when the agents start their activity and, thereby, can have only an awareness inherited in some way. In this case the governing fractional differential equation is shown to be of the Caputo type.

The dynamics of systems comprising two and three identical agents with the rock-paper-scissors interaction is analyzed in detail. First, the system stability is studied analytically and, then, the instability development is investigated numerically. In particular, it has been found that the longer memory, the more easily the Nash equilibrium looses stability, which is in agreement with the model of self-organization induced by dynamics of uncertainty \cite{RD2}. Roughly speaking the agents with weak memory cannot recognize the presence of other agents and treat their influence as random fluctuations in the environment.

For the two agent system the numerical analysis of the instability has demonstrated the existence of two its modes significantly different in their properties. One mode matches a limit cycle in the system phase space, stable periodic oscillations in the probability of agents choice, and a certain mismatch in the agent behavior. This mode undergoes the subcritical bifurcation and is dominant when the system parameters lie inside the instability region far enough from its boundary. The other mode undergoing the supercritical bifurcation describes oscillations in the agent preference whose amplitude and period grow continuously in time, at least, within the simplified model used in numerical simulation. These oscillations could be stabilized by the limit capacity of the agent memory, which however is worthy of individual investigations. The latter mode matches the synchronized behavior of the two agents and can be regarded as ``altruism self-organization''. The studied transient processes enable us to assume that there should be other modes of the system instability which, however, seem to be metastable. These phenomena are due to scale-free memory, at least, for actually the same model where, however, the memory is described by a single scale, only the first instability mode has been found and it undergoes the supercritical bifurcation.

In the three agent system the interaction of agents destroys the first mode and the system dynamics becomes irregular. However, near the instability threshold the second mode can survive and emerge after rather long and complex transient processes. For all the initial conditions randomly generated in numerical simulation the dynamics of the three agent system exhibits anomalous attraction to the equilibrium point, namely, it periodically returns to the equilibrium point being unstable and resides in its vicinity during remarkable time intervals. This behavior resembles the observed dynamics of trimorphic populations with evolutionary stable cycle variations alternated by breakdown events giving rise to dimorphic or monomorphic populations.

The obtained results enable us also to presume that to describe the observed complex phenomena in speciation a relevant model should deal with not only the proportions of species but also some variables quantifying the process of species individual learning in adapting to variations in the environment as well as the population structure.

\begin{acknowledgement}
The authors appreciate the support of RFBR Grants 09-01-00736 as well as the research support R-24-4 from the University of Aizu.
\end{acknowledgement}

\appendix

\section{Mean field approximation}\label{APP:MFA}

Let us consider a generalized form of the update rule~\eqref{mb:3a}
\begin{multline}\label{app:1}
   Q_{a}(x,t_{n+1})= W\big[P_a(x,t_n)\big]\cdot \delta_{xx_{a}} \Delta \, R_{a}(x|\mathcal{X}_{a})
\\
{}-\frac{\Delta}{T}\, Q_{a}(x,t_{n})+Q_{a}(x,t_{n})\,,
\end{multline}
where the weight coefficient $W\big[P_a(x,t_n)\big]$ is assumed, at first, to be a certain smooth function of the choice probability $P_a(x,t_n)$. The system update at the step $t_n$ is determined by the collection of expressions~\eqref{app:1}, where the agent index $a$ and the action index $x$ run over all their possible values independently of each other and the Kronecker delta  $\delta_{x,x_a}$ reflects the choice of the action $x_a$ by the agent $a$ at the time moment $t_n$.

According to the adopted assumption~\eqref{mb:4} the quantities $\{Q_a(x,t)\}$ cannot change substantially on scales of order $\Delta$. Therefore we can consider a composite step of the system update comprising $k$ elementary steps such that $1\ll k\ll T/\tau$ and to write for it the following equation
\begin{multline}\label{app:3}
   Q_{a}(x,t_{n+k})= W\big[P_a(x,t_n)\big]\Delta \sum_{x_a,\mathcal{X}_a\in \mathfrak{C}_{ak}} \delta_{xx_{a}} \, R_{a}(x|\mathcal{X}_{a})
\\
{}-\frac{k\Delta}{T}\, Q_{a}(x,t_{n})+Q_{a}(x,t_{n})\,.
\end{multline}
Here the symbol $\mathfrak{C}_{ak}$ stands for the set of actions
\begin{equation}\label{app:4}
    \mathfrak{C}_{ak} = \quad
    \begin{array}{c|cccc}
    {} & a_1 & a_2 & \dots & a_N\\\hline
    1 &\, x_{11} & x_{12} &\dots & x_{1N}\\
    2 &\, x_{21} & x_{22} &\dots & x_{2N}\\
    \dots& \hdotsfor{4}\\
    k &\, x_{k1} & x_{k2} &\dots & x_{kN}
     \end{array}
\end{equation}
where $x_{ij}$ is the action taken by the agent $a_j$ at the elementary step $t_{n+i}$ of the system update. Each row in table~\eqref{app:4} represents the set $x_a\bigcup\mathcal{X}_a$ of actions at the corresponding time from the standpoint of the agent $a$.

The updated values $Q_a(x,t_{n+k})$ are determined by a specific realization of the set $\mathfrak{C}_{ak}$ so, strictly speaking, they are random quantities.  However when the time capacity $T$ of the agent memory is high enough we can choose the number of elementary steps sufficiently large that any agent explore many options of its choice as well as the choice of the other agents. In this case the sum entering equation~\eqref{app:3} may be replaced by the sum running over all the possible realizations of the set $x_a\bigcup\mathcal{X}_a$ with the weights specified by their probabilities
\begin{equation}\label{app:5}
    \mathcal{P}\left(x_a\bigcup\mathcal{X}_a|t\right) = P_a(x_a,t)\prod_{a'\in\mathfrak{U}_a} P_{a'}(x_{a'},t)\,.
\end{equation}
It should be noted that it is exactly the point where the adopted assumption about the mutual independence of the agents in their actions has been taken into account.  The replacement
\begin{equation}\label{app:6}
    \sum_{x_a,\mathcal{X}_a\in \mathfrak{C}_{ak}}\ldots \Rightarrow
    k\sum_{x_a,\mathcal{X}_a} \mathcal{P}\left(x_a\bigcup\mathcal{X}_a|t\right)\ldots
\end{equation}
converts equation~\eqref{app:3} into the following
\begin{subequations}\label{app:7}
\begin{multline}\label{app:7a}
   Q_{a}(x,t_{n+k})= k\Delta\,\bigg\{W\big[P_a(x,t_n)\big] P_a(x,t_n)
\\
   {}\times\sum_{\mathcal{X}_a} R_{a}(x|\mathcal{X}_{a})
   \prod_{a'\in\mathfrak{U}_a} P_{a'}(x_{a'},t_n)
{}-\frac{1}{T}\, Q_{a}(x,t_{n}) \bigg\}
\\
    {} + Q_{a}(x,t_{n})
\end{multline}
and treating the quantities $\{Q_a(x,t_n)\}$ as continuous functions of time we get
\begin{multline}\label{app:7b}
   \frac{dQ_{a}(x,t)}{dt}= W\big[P_a(x,t)\big] P_a(x,t)
\\
   {}\times\sum_{\mathcal{X}_a} R_{a}(x|\mathcal{X}_{a})
   \prod_{a'\in\mathfrak{U}_a} P_{a'}(x_{a'},t)
    -\frac{1}{T}\, Q_{a}(x,t)\,,
\end{multline}
\end{subequations}
which is the desired mean field approximation of the update rule~\eqref{app:1}.

Keeping in mind the reasoning presented in Sec.~\ref{sec:2.1} we convert from the quantities $Q_a(x,t)$ to the quantities $q_a(x,t)$ showing their relative variations, i.e.,
\begin{equation}
\label{app:7add1}
     q_a(x,t) = Q_a(x,t) - \frac1M  \sum_{x'\in\mathfrak{X}} Q_a(x',t)\,.
\end{equation}
In these terms equation~\eqref{app:7b} reads
\begin{equation}\label{app:7add2}
     \frac{dq_a(x,t)}{dt} =  r_a(x,t)- \frac{1}{T}\, q_a(x,t)\,,
\end{equation}
where the relative reward rate is given by the expression
\begin{multline}
\label{app:7add3}
     r_a(x,t)  =  W\big[P_a(x,t)\big] P_a(x,t) \\
     {}\times \sum_{\mathcal{X}_a} R_a(x|\mathcal{X}_a) \prod_{a'\in\mathfrak{A}_a} P_{a'}(x_{a'},t)
\\
     {} - \frac1M \sum_{x'\in\mathfrak{X}}  W\big[P_a(x,t)\big] P_a(x,t)
\\
     {}\times\sum_{\mathcal{X}_a} R_a(x'|\mathcal{X}_a) \prod_{a'\in\mathfrak{A}_a} P_{a'}(x_{a'},t)\,.
\end{multline}
By virtue of relationship~\eqref{mb:1q} and expression~\eqref{app:7add3} the relative reward rate $r_a(x,t)$ is an autonomous function of the quantities $\{q_1(x,t),q_2(x,t),\ldots,q_N(x,t)\}$, thereby, the collection of equations~\eqref{app:7add2} forms a closed autonomous model for the reinforcement learning under consideration.

To complete the derivation of equation~\eqref{mb:5} we need to specify the $W(P)$-dependence.

\subsubsection*{The $W(P)$-ansatz}

Let us analyze the family of ans\"atze
\begin{equation}\label{app:0}
    W(P) = P^{-\nu}
\end{equation}
with a nonnegative exponent,  $\nu\geq 0$, which allows for possible overweighting of low-probability events. When the agents do not interfere with one another in their actions the payoff function $R_a(x|\mathcal{X}) = R_a(x)$ for any  agent $a$ does not depend on the choice of the other agents, thereby, the relative reward rate $r_a(x,t)$ in its list of arguments contains only the quantities $q_a(x)$. Let us consider in detail this situation when the set of possible actions consists of only two actions, $\mathfrak{X}=\{x_1,x_2\}$, which, in addition, are equivalent, i.e., $R_a(x_1) = R_a(x_2) = R$. Then using relationship~\eqref{mb:1q} the system of equations~\eqref{app:7add2} for a given agent $a$ is reduced to the equation
\begin{equation}\label{app:10}
    \frac{dq}{dt} = \frac{R}2\,\frac{\sinh\big((1-\nu)\beta q\big)}{[\cosh(\beta q)]^{1-\nu}} - \frac{q}{T}
\end{equation}
containing solely the variable $q=[Q_a(x_1)-Q_a(x_2)]/2$.

\begin{figure}
\begin{center}
\includegraphics[width=\columnwidth]{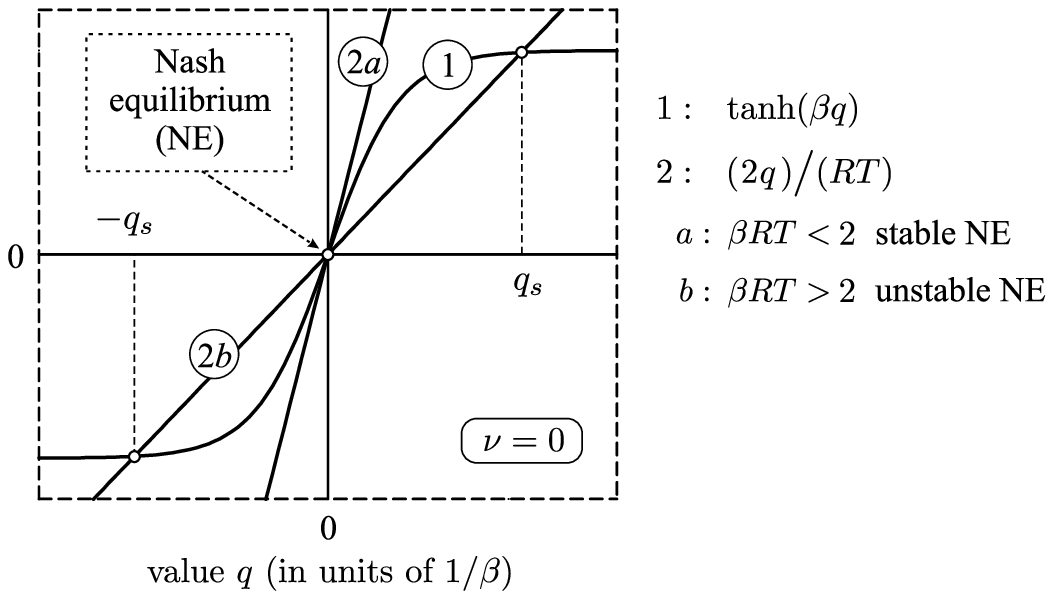}
\end{center}
\caption{Illustration of the instability mechanism for reinforcement learning governed by equation~\protect\eqref{app:10}.}
\label{Fig7}
\end{figure}

As it must be, the value $q=0$ matching the Nash equilibrium is a stationary point of equation~\eqref{app:10}. However in the limit of large values of the parameter $T$ the Nash equilibrium can lose the stability if $\nu <1$. Namely it is the case when the inequality
\begin{equation}\label{app:11}
(1-\nu)\beta RT > 2
\end{equation}
holds and equation~\eqref{app:10} admits three stationary points, the point $q = 0$ being now unstable and two new points $\pm q_s$ matching the stable dynamics of the given system in their vicinity. Figure~\ref{Fig7} illustrates this situation for $\nu=0$. This instability is an innate feature of the given algorithm; the learning processes does not lead to the hidden Nash equilibrium $q=0$ not due to some external perturbations induced, for example, by influence of the other agents but because of the fact that the given agent exploring more often those options that seem to it more preferable just faster accumulates their rewords.

To eliminate this mechanism of the Nash equilibrium instability from the analyzed model we adopt ansatz~\eqref{app:0} with $\nu = 1$.  In this case equation~\eqref{app:7add2} with expression~\eqref{app:7add3} immediately leads us to the governing equation~\eqref{mb:5} and express~\eqref{mb:6}.

\section{Equilibrium perturbations caused by random fluctuations in the reward rates and the system stability}\label{APP:B}

In order to analyze the impact of random fluctuations in the reward rate on the system dynamics we return to the integral equation~\eqref{mb:9final} and split the term $r_a(x,t)$ into two parts
\begin{equation}\label{app2:1}
    r_a(x,t) = r^r_a(x,t) + r^f_a(x,t)\,,
\end{equation}
where the former one is, as previously, the autonomous function of the action values, $r^r_a\big\{q_{a'}(x,t)\big\}$, specified by expression~\eqref{mb:6} and the latter summand represents small random fluctuations in the reward rate. Dealing with small perturbations in the system dynamics we may consider different components of the pattern $r^f_a(x,t)$ as well as the variations $\delta q_a(x,t)$ in the action values induced by these components separately. Therefore,  let us confine ourselves to the case when the random fluctuations $r^f_a(x,t)$ are located inside some internal interval $[t_1,t_2]$ (here $t_1> t_0$) and consider time scales such that $t-t_0 \gg t_2 - t_0$. Since no perturbations in $q_a(x,t)$ can occur before the time moment $t_1$ equation~\eqref{mb:9final} for $\delta q_a(x,t)$ and $t-t_0\gg t_2-t_0$ reads
\begin{multline}\label{app2:3}
     \delta q_a(x,t) = \tau^{1-\gamma}\int\limits_{t_0}^t dt'\,
     \frac{E_{\gamma,\gamma}\left[-\left(\tfrac{t-t'}{T}\right)^\gamma \right]}{(t-t')^{1-\gamma}}\, \delta r^r_a(x,t')
\\
     {}+ \frac{E_{\gamma,\gamma}\left[-\left(\tfrac{\vphantom{t'}t-t_0}{T}\right)^{\gamma}\right]}{(t-t_0)^{1-\gamma}} \int_{t_1}^{t_2} dt' r^f(x,t')\,,
\end{multline}
where $\delta r^r_a(x,t)$ is the regular component of the reward rate linearized with respect to small variations of $q_a(x,t)$ near the equilibrium point $q^\text{eq}_a(x)$.

Again the known relationship between the Cauchy type problems for factional differential equations and the Volter\-ra integral equations \cite{Kilbas} enables us to state that the asymptotics of the perturbations $\delta q_a(x,t)$ obeys the following fractional differential equation
\begin{equation}\label{app2:4}
   \widehat{D}_{t_0}^\gamma \delta q_a(x,t) = \tau^{1-\gamma} \delta r^r_a(x,t) - \frac1{T^{1-\gamma}}\,\delta q_a(x,t)\,,
\end{equation}
by virtue of \eqref{app2:3}. Here the left-hand side is the Riemann-Liouville fractional derivative of order $\gamma$ defined by the expression
\begin{equation}\label{app2:5}
   \widehat{D}_{t_0}^\gamma  q_a(x,t):=
   \frac1{\Gamma(1-\gamma)} \frac{d}{dt}\int\limits_{t_0}^t \frac{dt'}{(t-t')^{\gamma}} \, q_a(x,t')\,.
\end{equation}
The solution $\delta q_a(x,t)$ of equation~\eqref{app2:4} meets a certain initial integral condition that, in the frameworks of our analysis, can be replaced by to the requirement
\begin{equation}\label{app2:6}
   \lim_{t\to t_0}(t-t_0)^{1-\gamma} q_a(x,t) = C\,,
\end{equation}
where $C$ is some constant \cite{Kilbas}.

The eigenfunction of the Riemann-Liouville fractional derivative that meets requirement~\eqref{app2:6} and possesses the eigenvalue $\lambda$  is the so-called $\gamma$-exponential function \cite{Kilbas}
\begin{equation}\label{app2:7}
   e^{\lambda z}_\gamma := \frac1{(t-t_0)^{1-\gamma}}E_{\gamma,\gamma}\left[\lambda (t-t_0)^\gamma \right]
\end{equation}
with the asymptotic behavior as $t\to\infty$
\begin{subequations}\label{app2:asymp}
\begin{align}
\label{app2:asymp1}
    e_\gamma^{\lambda t} & = \frac{\lambda^{(1-\gamma)/\gamma}}{\gamma}
     e^{\left(\lambda^{1/{\gamma}}\,t\right)}
     + O\left(\frac1{t^{\gamma+1}}\right)
\\
\intertext{for $|\text{arg}(\lambda)|\leq \gamma\pi/2$ and}
\label{app2:asymp2}
      e_\gamma^{\lambda t} & =
    - \frac1{\lambda^2\Gamma(-\gamma)\cdot t^{\gamma+1}} + O\left(\frac1{t^{2\gamma+1}}\right)
\end{align}
\end{subequations}
for $\gamma\pi/2 <|\text{arg}(\lambda)|\leq \pi$.

The obtained results enable us to state the following. First, by virtue of \eqref{asymp} and \eqref{app2:asymp} the instability onset caused by perturbations in the initial conditions or random fluctuations in the reward rate matches the eigenvalues $\lambda$ of the corresponding fractional derivatives belonging to the same region on the complex plane. Second, equations~\eqref{mb:11} and \eqref{app2:4} governing the dynamics of perturbations induced by both of these mechanisms actually have the same form within the replacement ${}^C\!\widehat{D}_{t_0}^\gamma \Leftrightarrow \widehat{D}_{t_0}^\gamma $. Thereby the instability condition~\eqref{instab:1} found in Sec.~\ref{Sec:LSA} describes also the instability onset induced by the random fluctuations in the reward rate.

\end{document}